\documentclass[9pt]{article}
\usepackage{spconf,amsmath}
\usepackage{algorithm,algorithmic}
   \usepackage[pdftex]{graphicx}
  \graphicspath{{Figures/}}
\usepackage{multicol, blindtext}
\usepackage{amsmath}
\usepackage{enumerate}

\newcommand{\qed}{\nobreak \ifvmode \relax \else
      \ifdim\lastskip<1.5em \hskip-\lastskip
      \hskip1.5em plus0em minus0.5em \fi \nobreak
      \vrule height0.75em width0.5em depth0.25em\fi}

\begin{document}
\title{Regularized block Toeplitz covariance matrix estimation via Kronecker product expansions}

\name{{Kristjan Greenewald}, {Alfred O. Hero III}}

\address{{University of Michigan, Ann Arbor}}
%

%

\maketitle
\begin{abstract}

In this work we consider the estimation of spatio-temporal covariance matrices in the low sample non-Gaussian regime. We impose covariance structure in the form of a sum of Kronecker products decomposition \cite{tsiliArxiv,greenewaldArxiv} with diagonal correction \cite{greenewaldArxiv}, which we refer to as DC-KronPCA, in the estimation of multiframe covariance matrices. This paper extends the approaches of \cite{tsiliArxiv} in two directions. First, we modify the diagonally corrected method of \cite{greenewaldArxiv} to include a block Toeplitz constraint imposing temporal stationarity structure. Second, we improve the conditioning of the estimate in the very low sample regime by using Ledoit-Wolf type shrinkage regularization similar to \cite{chen2010shrinkage}. For improved robustness to heavy tailed distributions, we modify the KronPCA to incorporate robust shrinkage estimation \cite{chen2011robust}. Results of numerical simulations establish benefits in terms of estimation MSE when compared to previous methods. Finally, we apply our methods to a real-world network spatio-temporal anomaly detection problem and achieve superior results.

\end{abstract}

\section{Introduction}
\label{S:Intro}
Multivariate processes are ubiquitous in signal processing. Our goal is to learn correlations between (possibly non Gaussian) variables both within a time instant (frame) and across time. The covariance for multivariate temporal processes manifests itself as multiframe covariance, i.e., the covariance not only between variables in a single frame, but also between variables in a set of nearby frames. If each frame contains $p$ variables, then the covariance is described by a $pT$ by $pT$ matrix:
\begin{equation}
\mathbf{\Sigma}_t = \mathrm{Cov}\left[\{\mathbf{X_n}\}_{n=t-T}^{t-1}\right],
\end{equation}
where $\mathbf{X}_n$ denotes the $p$ variables in the $n$th frame. 

We focus on large $p$ small $n$ covariance estimation, that is, the high dimensional regime where the number of variables exceeds the number of training samples available to learn the covariance. For $p\geq n$, the standard sample covariance matrix (SCM) is the maximum likelihood covariance estimator. When $n$ is on the order of $p$ or smaller, however, it is well known that the SCM has a very undesirable poorly conditioned eigenstructure which results in the poor conditioning of the  inverse of the SCM, leading to poor estimates of the inverse covariance ($\mathbf{\Sigma}^{-1}$). Accurate estimation of the inverse covariance is critical for many applications, including graphical modeling, prediction, anomaly detection, and classification. 

One approach for addressing this problem is shrinkage estimation, which uses a weighted average of the sample covariance and a deterministic covariance, often chosen as a diagonal matrix called the shrinkage target. This can significantly improve the accuracy of the inverse of the estimate due to the improved eigenspectrum. The optimal amount of shrinkage, i.e., the shrinkage weights, is often estimated as the minimizer of the expected L2 estimation error \cite{ledoit2004well,chen2010shrinkage}. When the samples are heavy-tailed non-Gaussian an improved shrinkage estimator can be constructed by sphering the samples and using Tyler ML iterations, making the covariance estimate more robust to outliers \cite{chen2011robust}.

In this work, we develop shrinkage estimators of covariance matrices having spatiotemporal structure. Estimation of spatio-temporal covariance matrices based on reducing the number of parameters via a truncated sum of Kronecker products representation, which we call KronPCA, is discussed in \cite{tsiliArxiv,greenewaldArxiv}, and 
asymptotic performance analysis \cite{tsiliArxiv} predicts significant gains in estimator MSE in the high dimensional regime as $n,p \rightarrow \infty$. However, the empirical performance suffers from poor conditioning of the estimate.

To compensate for this poor conditioning of KronPCA we introduced diagonally corrected Kronecker PCA (DC-KronPCA) \cite{greenewaldArxiv}. This paper extends \cite{greenewaldArxiv} in two directions. First, the MSE analyses of the shrinkage estimator in \cite{chen2010shrinkage,chen2011robust} is used to estimate the shrinkage parameter of DC-KronPCA in the Gaussian and heavy-tailed regimes, respectively. Second, when the measurements are known to be temporally stationary, DC-KronPCA is formulated with a temporal block Toeplitz constraint \cite{wiesel2013time}. 

To illustrate the proposed DC-KronPCA based methods, we consider a real world application: spatiotemporal anomaly detection in sensor networks. We incorporate temporal information into the spatio-temporal covariance as opposed to the single frame covariance based method of \cite{chen2011robust}. 

The rest of this paper is organized as follows: in Section \ref{S:Kron}, we review DC-KronPCA and present our block Toeplitz modification of it. Our methods for standard and robust shrinkage of the KronPCA estimate are given in Section \ref{S:Shrink}, and numerical simulation results are presented in Section \ref{S:SimRes}. We discuss an application to sensor network anomaly detection in Section \ref{S:Anom}, and conclude in Section \ref{S:Conc}.

\section{Block Toeplitz Kronecker PCA}
\label{S:Kron}
In this section, we consider the regularization of the sample covariance via decomposition into a sums of (space vs. time) Kronecker products representation (KronPCA) \cite{tsiliArxiv,greenewaldArxiv}. 
We thus develop a method for learning block Toeplitz KronPCA covariances.

Let $\mathbf X$ be a $p \times T$ matrix with entries $\tilde{x}(m,t)$ denoting samples of a space-time random process defined over a $p$-grid of space samples $m\in \{1,\ldots, p\}$ and a $T$-grid of time samples $t\in \{1,\ldots, T\}$. Let ${x}={\mathrm{ vec}}({\mathbf{X}})$ denote the $pT$ column vector obtained by lexicographical reordering. Define the $pT \times pT$ spatiotemporal covariance matrix $\mathbf \Sigma=\mathrm{Cov}[{x}]$. 
Following \cite{greenewaldArxiv,tsiliArxiv,kamm2000optimal}, we define the DC-KronPCA model
\begin{equation}
\label{SumApprox}
\mathbf{{\Sigma}} = \left(\sum\nolimits_{i = 1}^{r}\mathbf{T}_i \otimes \mathbf{S}_i\right) + \mathbf{I} \otimes \mathbf{U},
\end{equation}
where $\mathbf T_i$ are $T\times T$ Toeplitz matrices (the temporal Kronecker factors), $\mathbf{S}_i$ are $p\times p$ matrices (the spatial Kronecker factors), $\mathbf{U}$ is a $p\times p$ diagonal matrix.

Following the approach of \cite{tsiliArxiv}, we propose to fit the model \eqref{SumApprox} to the sample covariance matrix ${\mathbf \Sigma}_{SCM}=n^{-1}\sum_{i=1}^n (x_i-\overline{x}) (x_i-\overline{x})^T$, where $\overline{x}$ is the sample mean, and $n$ is the number of samples of the space time process $\mathbf X$. The estimator of the parameters $\mathbf T_i$, $\mathbf S_i$ and $\mathbf U$ in \eqref{SumApprox} is performed by minimizing the following objective function
\begin{equation}
\|\mathbf R-\hat{\mathbf R}\|_F^2+\beta\|\hat{\mathbf R} - \mathcal{R}(\mathbf{I}\otimes \mathbf{U})\|_*,
\end{equation}
where $\mathbf R=\mathcal{R}(\mathbf {\Sigma}_{SCM})$ and $\mathcal{R}$  denotes the permutation rearrangement operator defined in \cite{tsiliArxiv,werner2008estimation} which maps $pT\times pT$ matrices to $T^2 \times p^2$ matrices. The objective function is minimized over all $T^2 \times p^2$ matrices $\hat{\mathbf R}$ that satisfy the constraint that $\mathcal { R}^{-1}(\hat{\mathbf{ R}})$  is of the form \eqref{SumApprox}.



This is equivalent \cite{greenewaldArxiv} to $\hat{\mathbf{\Sigma}} = \mathbf{I} \otimes \mathbf{U} + \mathcal{R}^{-1}(\hat{\mathbf{R}})$ where $\mathcal R^{-1}$ is the depermutation operator (inverse of $\mathcal R$) and $\hat{\mathbf{R}}$ is found by solving
\begin{align}
\min_{\hat{\mathbf{R}}}& || \mathbf{M} \circ(\mathbf{R} - \hat{\mathbf{R}}) ||_F^2 + \beta \| \hat{\mathbf{R}} \|_*,\\\nonumber
\hat{\mathbf{R}} &= \sum_{i = 1}^r t_i s_i^T, \:
s.t. \: \mathbf{T}_i \: \mathrm{Toeplitz}, \: \forall i,
\end{align}
where the $t_i,s_i$ are $\mathbf{T}_i, \mathbf{S}_i$ permuted as in \cite{werner2008estimation,greenewaldArxiv}, $\mathbf{M}$ is a matrix masking out the elements corresponding to the covariance diagonal, and $\circ$ denotes the elementwise product).
Following the results of \cite{kamm2000optimal}, we note that the Toeplitz constraint on $\mathbf{T}_i$ is equivalent to
\begin{equation}
\label{Eq:Toeplitz}
[t_i]_{k} = a^{(i)}_{j + T},\: \forall k \in \mathcal{K}(j), j \in [-T+1,\: T-1],
\end{equation}
for some vector $a^{(i)}$ where
\begin{equation}
\mathcal{K}(j) = \{k : \: (k-1)T + k + j \in [-T+1,\: T-1]\}.
\end{equation}
Let
\begin{equation}
\label{Eq:Tweight}
\tilde{t}^{(i)}_{j+T}= a^{(i)}_{j + T}\sqrt{T - |j|}, \quad j \in [-T+1,\: T-1].
\end{equation}

Following the method of \cite{kamm2000optimal}, the optimization can be shown to be equivalent to
\begin{align}
\label{Eq:OptProb}
\min_{{\tilde{\mathbf{R}}}} ||\tilde{\mathbf{M}} \circ(\mathbf{B} - {\tilde{\mathbf{R}}}) ||_F^2 + \beta \| \tilde{\mathbf{R}} \|_*,\:
{\tilde{\mathbf{R}}} = \sum_{i = 1}^r \tilde{t}_i s_i^T, \:
\end{align}
where $\tilde{\mathbf{R}} = \mathcal{P}(\hat{\mathbf{R}})$, $\tilde{\mathbf{M}} = \textrm{sign}(\mathcal{P}(\mathbf{M}))$, and $\mathbf{B} = \mathcal{P}(\mathbf{R})$. The operator $\mathcal{P}$ (from $T^2\times p^2$ to $(2T-1) \times p^2$ matrices) is defined as $\tilde{\mathbf{A}} = \mathcal{P}(\mathbf{A})$ such that
\begin{equation}
\tilde{A}_{j+T} = \frac{1}{\sqrt{T - |j|}}\sum_{k \in \mathcal{K}(j)}A_k,  \quad \forall j \in [-T+1, \: T-1],
\end{equation}
where $A_j$ is the $j$th row of $\mathbf{A}$. Note that this corresponds to an unbiased Toeplitz estimate. It is of interest to investigate the extension of biased Toeplitz estimators \cite{vinogradova2014estimation} to this case.

This well-studied optimization problem (nuclear norm penalized low rank matrix approximation with missing entries) is considered in \cite{mazumder2010spectral}, where it is shown to be convex. Several fast solution methods are available, including the iterative SVD-based method of \cite{mazumder2010spectral}.
If diagonal correction is not used (i.e. $\mathbf{M}$ is all ones), the resulting optimization problem can be solved directly using the SVD \cite{tsiliArxiv}. The block Toeplitz diagonally corrected covariance estimate is given by
\begin{align}
\label{Eq:KronPCA}
\hat{\mathbf{\Sigma}} = \mathbf{I} \otimes \mathbf{U}+\mathcal R^{-1}\left(\mathcal{P}^{*} \left( \tilde{\mathbf{R}} \right)\right),
\end{align}
where $\tilde{\mathbf{R}}$ is the minimizer of \eqref{Eq:OptProb} and $\mathcal P^{*}$ is defined by $\mathbf{A} = \mathcal{P}^{*}(\tilde{\mathbf{A}})$ where
\begin{equation}
{A}_{k} = \frac{1}{\sqrt{T - |j|}}\tilde{A}_{j+T},  \: \forall k \in \mathcal{K}(j),\:\forall j \in [-T+1, \: T-1].
\end{equation}
The diagonal matrix $\mathbf{U}$ is set in the next section.

\section{Kronecker PCA Shrinkage Estimation}
\label{S:Shrink}

\subsection{Diagonal Shrinkage}
Diagonal shrinkage shrinks the sample covariance towards a scaled identity matrix. This improves the conditioning of the estimate, which makes the inverse of the estimate more stable.
\begin{equation}
\hat{\mathbf{\Sigma}} = (1-\hat{\rho})\hat{\mathbf{\Sigma}}_{kron} + \hat{\rho}\mathbf{F},
\end{equation}
where $\mathbf{F} = \frac{\mathrm{trace}(\hat{\mathbf{\Sigma}}_{kron})}{pT}\mathbf{I}$ and $\hat{\mathbf{\Sigma}}_{kron}$ is the block Toeplitz DC-KronPCA estimate of the covariance. 

%

It remains to determine the amount of shrinkage, i.e. $\hat{\rho}$. We use the Ledoit-Wolf (LW) \cite{ledoit2004well} solution that asymptotically (large sample size $n$)
minimizes Frobenius estimation error when $\hat{\mathbf{\Sigma}}$ is the SCM. Due to the lower variance of the DC-KronPCA covariance estimate relative to the SCM, the amount of shrinkage is expected to be overestimated.

Since the LW shrinkage solution minimizes the estimator MSE it depends on the true (but unavailable) covariance, as in \cite{ledoit2004well,chen2010shrinkage} we use a plug-in estimator of $\hat{\rho}$ that uses the DC-KronPCA estimator denoted $\hat{\mathbf{\Sigma}}_{kron}$, given by \eqref{Eq:KronPCA}. We call this method DC-KronPCA-LW.

\subsection{Robust Shrinkage}
LW methods are not designed to be robust to heavy tailed distributions \cite{chen2011robust}. We propose to modify the iterative algorithm of \cite{chen2011robust}, which in turn is a modification of the Tyler iterations for ML estimation of the covariances of elliptical distributions \cite{chen2011robust} incorporating shrinkage at the end of each iteration.



Define the operator $\mathrm{KronPCA}_{\mathbf{T}}\{\cdot\}$ that returns the Toeplitz (temporal) Kronecker factor resulting from running block Toeplitz KronPCA with $r=1$. We apply the Chen-Tyler iterations \cite{chen2011robust} modified to incorporate a projection onto covariances of the form $\mathbf{T}\otimes \mathbf{S}$. In order to guarantee positive semidefiniteness of the projection, we optimize the ML KronPCA objective function with the Toeplitz Kronecker factor held constant \cite{werner2008estimation}. This problem is convex and has a closed form solution given by \cite{werner2008estimation}. Our robust DC-KronPCA estimation algorithm is detailed in Algorithm \ref{Alg:DiagKron}.

Following the same rationale as in the previous subsection, we use the solution of \cite{chen2011robust} for $\hat{\rho}$, which aims to minimize the expected MSE for the SCM based method. For the plug-in covariance estimate needed in the solution \cite{chen2011robust}, we use the DC-KronPCA-LW estimate. 

\section{Simulation Results}
\label{S:SimRes}
\begin{figure}[htb]
\centering
\includegraphics[width=2.9in]{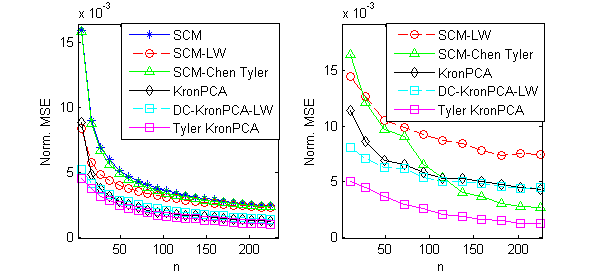}
\caption{Estimation of Kronecker product covariance from i.i.d. samples. Left: Normalized MSE as a function of $n$ for Gaussian samples. Right: Normalized MSE for heavy tailed multivariate-t samples. Note the universal superiority of the Kronecker methods, the low sample improvement resulting from DC-KronPCA-LW, and the superiority of the robust shrinkage KronPCA in the heavy tailed case.}
\label{Fig:Gauss}
\end{figure}

We conduct numerical simulations to evaluate the relative MSE performance of the various covariance estimation algorithms considered. We first consider learning a $p=100,\:t = 10$ multiframe covariance defined as the Kronecker product of the covariances of two AR(1) processes, one across time (with AR coefficient .5) and the other space (AR coefficient .95). For various values of $n$, we generate $n$ iid samples, learn the covariance, and compute the MSE. The methods compared are the SCM, the Ledoit Wolf estimator (SCM-LW), KronPCA, DC-KronPCA-LW, the robust shrinkage estimator of \cite{chen2011robust} (Chen Tyler), and our robust KronPCA shrinkage (Tyler-KronPCA). For this experiment, the Kronecker methods use $r=1$, i.e. a single Kronecker summand in \eqref{SumApprox}. The results for Gaussian measurements $\mathbf X$  are shown in Figure \ref{Fig:Gauss} (left), with all performance curves averaged over 100 trials. Figure \ref{Fig:Gauss} (right) shows results (averaged over 3000 trials) for the heavy tailed elliptical multivariate-t distribution given by multiplication of the AR multivariate Gaussian by $\sqrt{d/\chi_d^2}$, where $d=3$ and $\chi_d^2$ is chi squared distributed.

\begin{algorithm}[H]

\caption{Robust DC-KronPCA Shrinkage Estimation}
\label{Alg:DiagKron}
\begin{algorithmic}[1]
\STATE $s_i = \frac{x_i}{\|x_i\|_2}$, $\forall i$.
\STATE Run Chen iterations until convergence, obtain $\hat{\mathbf{\Sigma}}$.
\STATE $\tilde{\mathbf{\Sigma}} \gets \hat{\mathbf{\Sigma}}$.
\WHILE{not converged}
\STATE $\hat{\mathbf{T}} = \mathrm{KronPCA}_{\mathbf{T}}\{\tilde{\mathbf{\Sigma}}\}$.
\WHILE{not converged}
\STATE Run Tyler iteration: $\tilde{\mathbf{\Sigma}} \gets \frac{pT}{n}\sum_{i=1}^n \frac{s_i s_i^T}{s_i^T \hat{\mathbf{\Sigma}}^{-1} s_i}$
\STATE $\hat{\mathbf{S}} = \frac{1}{T}\sum_{i,j = 1}^{T} [\hat{\mathbf{T}}^{-1}]_{ij}\tilde{\mathbf{\Sigma}}_{ji}$.
\STATE $\hat{\mathbf{\Sigma}} \gets \hat{\mathbf{T}} \otimes \hat{\mathbf{S}}$
\STATE $\hat{\mathbf{\Sigma}} \gets (1-\rho)\frac{pT}{\mathrm{trace}(\hat{\mathbf{\Sigma}})}\hat{\mathbf{\Sigma}} + \rho \mathbf{I}$
\ENDWHILE

\ENDWHILE
\STATE Return $\hat{\mathbf{\Sigma}}$.
\end{algorithmic}
\end{algorithm}

\section{Application to Anomaly Detection in Wireless Sensor Networks}
\label{S:Anom}

In this section we apply our proposed spatio-temporal covariance estimators to activity detection in a
wireless sensor network consisting of 14 transmit/receive Mica2 sensor nodes placed randomly in a lab at the University of Michigan \cite{chen2011robust}.
All sensors transmitted asynchronously every 0.5 seconds over a period of 30 minutes, and the received signal strength (RSS) of every transmission was recorded by the remaining sensors, giving $182$ measurements. Disturbances in the RSS values occurred when people passed through the lab at various times during this period. Ground truth was recorded by a camera \cite{chen2011robust}. The dataset is detailed in \cite{chen2011robust}.
As in \cite{chen2011robust}, the data was detrended, the covariance learned from a subset of the data, and anomaly detection performed by thresholding the Mahanolobis distance. 

For detecting anomalies in time series data, we divide the sequence up into overlapping chunks of length $T$ using a sliding window and determine whether each one is anomalous independently. Chunks are included in testing only if they are $>75$\% anomalous or nonanomalous. 

\begin{figure}[htb]
\centering
\includegraphics[width=1.8in]{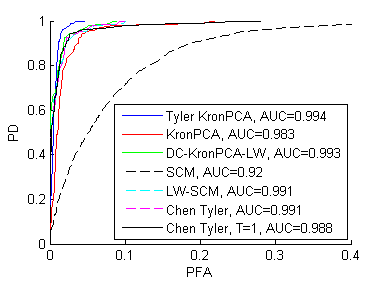}\includegraphics[width=1.8in]{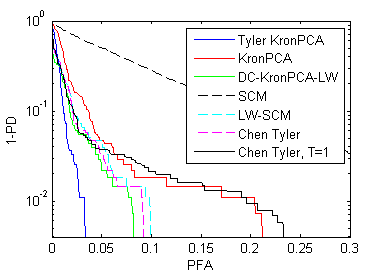}
\caption{RSS network anomaly ROC curves for various covariance estimation methods with $T=10$, with the corresponding areas under the curve (AUC). Note the superiority of the Kronecker-based shrinkage methods.}
\label{Fig:Fixed}
\end{figure}

\begin{figure}[htb]
\centering
\includegraphics[width = 3.6in]{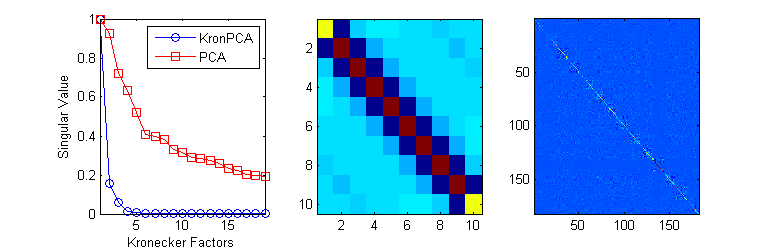}
\caption{Left: Normalized RMS values of (block Toeplitz) Kronecker components and first few PCA components for the SCM learned from the RSS dataset. Note that fewer KronPCA than PCA components are required to achieve a given representation accuracy. Also shown are $\hat{\mathbf{T}}^{-1}$ (middle) and $\hat{\mathbf{S}}^{-1}$ (right), from robust DC-KronPCA shrinkage, $n = 200$.}
\label{Fig:Skree}
\end{figure}

The different covariance estimators described in Section \ref{S:SimRes} are used for anomaly detection. The results for the case of learning with $T=10$ on a fixed set of 200 consecutive nonanomalous frames are shown in Figure \ref{Fig:Fixed}. The multiframe advantage is clearly evident (compare line labeled Chen-Tyler, $T=1$ to the other curves). 
In Figure \ref{Fig:Skree}, we show the magnitudes of the 19 block Toeplitz Kronecker components of the $T=10$ SCM using the whole dataset ($n = 3000$). That KronPCA achieves higher concentration (fewer terms) than ordinary PCA is evident. We also show the learned Kronecker factors arising from robust DC-KronPCA shrinkage (for Figure \ref{Fig:Fixed}), inverted to show partial correlation structure.

%


\section{Conclusion}
\label{S:Conc}

In this work we considered the estimation of spatiotemporal covariances and their inverses in the low sample regime. We proposed a combination of Ledoit-Wolf type shrinkage techniques, KronPCA, and block Toeplitz structure for temporally stationary processes. In addition, we presented a modified version of DC-KronPCA exploiting the block Toeplitz structure implied by stationarity. We validated the performance advantages of the proposed covariance estimation strategies by running several simulations and experiments.

\section{Acknowledgements}
This work was partly supported by ARO under grant W911NF-11-1-0391 and AFRL under FA8650-07-D-1220-0006.

\bibliographystyle{IEEEbib}
\bibliography{CAMSAP_bib}

\begin{thebibliography}{10}

\bibitem{tsiliArxiv}
T.~Tsiligkaridis and A.~Hero,
\newblock ``Covariance estimation in high dimensions via kronecker product
  expansions,''
\newblock {\em IEEE Trans. on Sig. Proc.}, vol. 61, no. 21, pp. 5347--5360,
  2013.

\bibitem{greenewaldArxiv}
K.~Greenewald, T.~Tsiligkaridis, and A.~Hero,
\newblock ``Kronecker sum decompositions of space-time data,''
\newblock in {\em IEEE CAMSAP}, 2013.

\bibitem{chen2010shrinkage}
Y.~Chen, A.~Wiesel, Y.~Eldar, and A.~Hero,
\newblock ``Shrinkage algorithms for mmse covariance est.,''
\newblock {\em IEEE Trans. on Sig. Proc.}, vol. 58, no. 10, pp. 5016--5029,
  2010.

\bibitem{chen2011robust}
Y.~Chen, A.~Wiesel, and A.~Hero,
\newblock ``Robust shrinkage estimation of high-dimensional covariance
  matrices,''
\newblock {\em IEEE Trans. on Sig. Proc.}, vol. 59, no. 9, pp. 4097--4107,
  2011.

\bibitem{ledoit2004well}
O.~Ledoit and M.~Wolf,
\newblock ``A well-conditioned estimator for large-dimensional covariance
  matrices,''
\newblock {\em Journal of multivariate analysis}, vol. 88, no. 2, pp. 365--411,
  2004.

\bibitem{wiesel2013time}
A.~Wiesel, O.~Bibi, and A.~Globerson,
\newblock ``Time varying arma models for covariance estimation,''
\newblock {\em IEEE Trans. on Sig. Proc.}, vol. 61, no. 11, pp. 2791--2801,
  2013.

\bibitem{kamm2000optimal}
J.~Kamm and J.~Nagy,
\newblock ``Opt. kronecker product approx. of block toeplitz matrices,''
\newblock {\em SIAM Journal on Matrix Analysis and App.}, vol. 22, no. 1, pp.
  155--172, 2000.

\bibitem{werner2008estimation}
K.~Werner, M.~Jansson, and P.~Stoica,
\newblock ``On estimation of cov. matrices with kronecker product structure,''
\newblock {\em IEEE Trans. on Sig. Proc.}, vol. 56, no. 2, pp. 478--491, 2008.

\bibitem{vinogradova2014estimation}
J.~Vinogradova, R.~Couillet, and W.~Hachem,
\newblock ``Estimation of toeplitz covariance matrices in large dimensional
  regime with application to source detection,''
\newblock {\em arXiv preprint arXiv:1403.1243}, 2014.

\bibitem{mazumder2010spectral}
R.~Mazumder, T.~Hastie, and R.~Tibshirani,
\newblock ``Spectral regularization algorithms for learning large incomplete
  matrices,''
\newblock {\em Journal of Machine Learning Research}, vol. 11, pp. 2287--2322,
  2010.

\end{thebibliography}

\end{document}